\documentclass[12pt]{article}
\usepackage{epsf}

\usepackage{epsfig,graphics}
\usepackage {graphicx}
\usepackage {epsfig}
\usepackage {subfigure}
\usepackage {tabularx} 
\usepackage{rotate}	
\usepackage{slashed}

\usepackage{amsmath}
\usepackage{amsfonts}
\usepackage{amssymb}
\usepackage{graphicx}
\usepackage{cite}
\usepackage{color}

\usepackage{fancyhdr}
\usepackage{hyperref}

\newcommand{\bmat}{\left(\begin{array}}
\newcommand{\emat}{\end{array}\right)}

\def\yzero{\smash{\hbox{$y\kern-4pt\raise1pt\hbox{${}^\circ$}$}}}

\def\beq{\begin{equation}}
\def\eeq{\end{equation}}
\def\beqa{\begin{eqnarray}}
\def\eeqa{\end{eqnarray}}

\def\-{\hphantom{-}}

\def\s2{\frac{1}{\sqrt2}}

\def\beq{\begin{equation}}
\def\eeq{\end{equation}}
\def\beqa{\begin{eqnarray}}
\def\eeqa{\end{eqnarray}}

\def\IF{\relax{\rm I\kern-.18em F}}
\def\II{\relax{\rm I\kern-.18em I}}
\def\IP{\relax{\rm I\kern-.18em P}}
\def\IC{\relax\hbox{\kern.25em$\inbar\kern-.3em{\rm C}$}}
\def\IR{\relax{\rm I\kern-.18em R}}

\def\Dsl{\,\raise.15ex\hbox{/}\mkern-13.5mu D} %this one can be subscripted
\def\IZ{Z\kern-.4em  Z}

\catcode`\@=11   
\newdimen\@rotdimen
\newbox\@rotbox  

\def\@vspec#1{\special{ps:#1}}%  passes #1 verbatim to the output
\def\@rotstart#1{\@vspec{gsave currentpoint currentpoint translate
   #1 neg exch neg exch translate}}% #1 can be any origin-fixing transformation
\def\@rotfinish{\@vspec{currentpoint grestore moveto}}% gets back in synch 
\def\@rotr#1{\@rotdimen=\ht#1\advance\@rotdimen by\dp#1%
   \hbox to\@rotdimen{\hskip\ht#1\vbox to\wd#1{\@rotstart{90 rotate}%
   \box#1\vss}\hss}\@rotfinish}
\def\@rotl#1{\@rotdimen=\ht#1\advance\@rotdimen by\dp#1%
   \hbox to\@rotdimen{\vbox to\wd#1{\vskip\wd#1\@rotstart{270 rotate}%
   \box#1\vss}\hss}\@rotfinish}%
\def\@rotu#1{\@rotdimen=\ht#1\advance\@rotdimen by\dp#1%
   \hbox to\wd#1{\hskip\wd#1\vbox to\@rotdimen{\vskip\@rotdimen
   \@rotstart{-1 dup scale}\box#1\vss}\hss}\@rotfinish}%
\def\@rotf#1{\hbox to\wd#1{\hskip\wd#1\@rotstart{-1 1 scale}%
   \box#1\hss}\@rotfinish}%
\def\rotate{\@ifnextchar[{\@rotate}{\@rotate[l]}}
\def\@rotate[#1]#2{\setbox\@rotbox=\hbox{#2}\@nameuse{@rot#1}\@rotbox}

\catcode`\@=12
\topmargin
-1.5cm
\textwidth
15.5cm
\textheight
23.5cm
\oddsidemargin
0.7cm
\evensidemargin
0.7cm

\begin{document}

%----------------------------------------------------------------------%
%  numbering equations with section number
%----------------------------------------------------------------------%
\makeatletter
\@addtoreset{equation}{section}
\makeatother
\renewcommand{\theequation}{\thesection.\arabic{equation}}
%----------------------------------------------------------------------%
%  title page
%----------------------------------------------------------------------%
\pagestyle{empty}

\vspace{-0.2cm}
\rightline{ IFT-UAM/CSIC-17-081}

\vspace{1.2cm}
\begin{center}

\LARGE{ A Note on the WGC, Effective Field \\ 
Theory and Clockwork within String Theory \\ [13mm]}

  \large{Luis E. Ib\'a\~nez${}^1$ and Miguel Montero${}^2$ \\[6mm]}
\small{
  ${}^1$ Departamento de F\'{\i}sica Te\'orica
and Instituto de F\'{\i}sica Te\'orica UAM/CSIC,\\[-0.3em]
Universidad Aut\'onoma de Madrid,
Cantoblanco, 28049 Madrid, Spain
\\ ${}^2$Institute for Theoretical Physics and\\
Center for Extreme Matter and Emergent Phenomena,\\
Utrecht University, Princetonplein 5, 3584 CC Utrecht, The Netherlands
\\[8mm]}
\small{\bf Abstract} \\[6mm]
\end{center}
\begin{center}
\begin{minipage}[h]{15.22cm}
It has been recently argued that Higgsing of theories with $U(1)^n$ gauge interactions  consistent with
the  Weak Gravity Conjecture (WGC)  may lead to effective field theories parametrically violating WGC constraints. The minimal examples typically
involve Higgs scalars with a large charge with respect to a $U(1)$ (e.g. charges  $(Z,1)$ in $U(1)^2$ with $Z\gg 1$). 
This type of Higgs multiplets play also a key role in clockwork $U(1)$ theories. We study these issues in the
context of  heterotic string theory and find that, while indeed there is no new physics at the standard magnetic WGC scale $\Lambda\sim g_{IR} M_P$, the string scale is just slightly above, at a scale $\sim \sqrt{k_{IR}}\Lambda$. Here $k_{IR}$ is the
 level of the IR $U(1)$ worldsheet current. We show that, unlike the standard magnetic cutoff, this bound is insensitive to subsequent Higgsing. One may argue that this constraint gives rise to no bound at 
 the effective field theory level since $k_0$ is model dependent and in general unknown. However there is an additional constraint 
 to be taken  into account, which is that the Higgsing scalars with large charge $Z$ should be part of the string massless spectrum,
 which becomes an upper bound $k_{IR}\leq k_0^2$, where $k_0$ is the level of the UV currents. Thus, for fixed $k_0$, $Z$ cannot be made parametrically large.
 The  upper bound on the charges $Z$ leads to limitations on the size and structure of hierarchies in an iterated
 $U(1)$ clockwork mechanism.

\end{minipage}
\end{center}
\newpage
%----------------------------------------------------------------------%
%  Resetting of counters
%----------------------------------------------------------------------%
\setcounter{page}{1}
\pagestyle{plain}
\renewcommand{\thefootnote}{\arabic{footnote}}
\setcounter{footnote}{0}
%----------------------------------------------------------------------%
%  Paper begins
%----------------------------------------------------------------------%

%\end{document}

%\end{document}

\tableofcontents

%&&&&&&&&&&&&&&&&&&&&&&&&&&&&&&&&&
\section{Higgsing and WGC constraints}

There has been recently a revived activity in trying to characterize theories which cannot possibly be embedded 
into a consitent theory of quantum gravity (and hence belong to the {\it swampland} \cite{WGC}) from those 
which can, see \cite{WGC1,WGC2} for a partial list of recent references. The classical constraint is the 
Weak Gravity Conjecture (WGC) \cite{WGC}  which, as applied to a $U(1)$ theory, states that a particle with charge $q$
and mass
\beq
m^2\ \leq q^2g^2M_p^2 \ ,
\label{wgc1}
\eeq
must exist in the low energy spectrum. A {\it strong}  version of the conjecture states that the lightest charged  state in the
theory must obey the inequality.
The magnetic version of the constraint indicates there must be an associated
scale of new physics at
\beq
\Lambda^2 \ \simeq \ g^2 M_p^2 \ .
\label{magwgc}
\eeq
So making $g\ll 1$ is not an innocent weakly coupled limit but implies also a new physics scale.  The strongest support for 
the WGC comes from not having found any counterexamples in string theory, although  no formal proof exists at the moment.
This type of bounds may be generalized from particles
to instantons, domain walls and branes leading to interesting constraints on axion cosmology, relaxions and other interesting physical systems
\cite{WGC,WGC1,WGC2}.

An important question is whether one can identify effective field theories which can be ruled out on the basis of the WGC. In particular 
if we find an effective low-energy theory violating eq.(\ref{wgc1}) can we conclude that the theory is in the swampland? 
In ref.\cite{saraswat} it was claimed that the answer to this question is {\it no} and an explicit example was shown in which, starting from
a theory obeying the WGC, Higgsing can lead to an effective theory parametrically violating eq.(\ref{wgc1}). So from a low energy analysis 
we cannot conclude whether a theory is consistent or not.  It is interesting to review this simple model before we go to its string theory version. 
There is a $U(1)_1\times U(1)_2$ theory with gauge couplings $g_{1,2}$ and particles  $\psi_{1,2}$ with charges $(1,0)$ and $(0,1)$ respectively.
We assume they have masses $m_{1,2}$ consistent with a bound as in eq.(\ref{wgc1}).
Consider now a  massless scalar with charges $(Z,1)$ getting a vev $v$.  The gauge symmetry is broken like 
$U(1)_1\times U(1)_2\rightarrow U(1)$ with a massless gauge linear combination (we take for simplicity $g_{1,2}=g$ and $Z\gg 1$ here)
\beq 
A\ =\ A_2 \ - \ \frac {1}{Z} A_1 \ = \left(-\frac{1}{Z},1\right)\cdot \vec{A}.
\eeq
The latter equality shows that $A$ has norm $1$, modulo $1/Z^2$ corrections. 
Now the field $\psi_1$ couples to  the massless gauge boson with an effective charge $q_1=-1/Z$.  One can make in principle $Z$ parametrically
large,  so that the WGC bound eq.(\ref{wgc1}) for $A$  is violated for $\psi_1$, even if originally the WGC bound before Higgsing is satisfied, as long as the mass $m_1$ satisfies $(g/Z)^2M_P^2< m_1^2<g^2M_P^2$. One would also predict  (wrongly) the existence of  new physics at a scale $\simeq (g/Z) M_p$.  Thus this is an example of how a perfectly consistent
theory is inconsistent with the standard strong and magnetic forms of the WGC upon Higgsing. As we will see shortly, in a string context we can do better. 

\section{A String Theory Embedding}

Let us consider now a string theory 4D compactification with some $U(1)^N$ gauge group and charged matter fields.
We consider here heterotic compactifications, which are particularly versatile in producing massless spectra with a
variety of charges, sometimes quite large. We will not need to specify a  construction and our arguments will apply to any kind of
compactification (CY, Gepner, Orbifold, etc) with or without supersymmetry. In heterotic compactifications with any arbitrary 
non-Abelian $G$  or Abelian group  $U(1)^N$ the gauge and gravity interactions are unified as
(we use conventions as in \cite{BOOK})
\beq
\alpha_Gk_G \ =\  \alpha_i k_i \ =\  16\frac {M_S^2}{M_p^2} \ ,
\label{unif}
\eeq 
where $M_S^2=1/\alpha '$, $\alpha_G,\alpha_i$ are the corresponding fine-structure constants and $k_G$ is the Kac-Moody level associated to
each non-Abelian gauge factor (a positive integer). In the simplest models $k_G=1$ whereas  for  $U(1)$'s $k_i$ is a normalization factor of each Abelian factor 
which depends on the choice of $U(1)$ generator. It is a positive rational number  (integer if all particle charges are chosen integer) which in principle may be arbitrarily large
\footnote{E.g., in models of particle physics one would like to have $k_Y/k_3=k_Y/k_2= 5/3$ to get canonical normalization for the
hypercharge $Y$. If the assumption of quantized charges is dropped, $k_1$ may be in some specific cases relatively large,
see e.g. \cite{Chaudhuri:1995ve} for examples with $k_1\sim 21$.}.   To be more precise,  with the world-sheet current for a $U(1)$ defined as $J_Q=iQ_I\partial X_I$, with OPE expansion
\beq
J_Q(z)J_Q(w)\ \simeq \  \frac {Q^2}{(z-w)^2}\ +\ ...
\eeq
one has $k_1=2Q^2$. Here $X_I, I=1-16$ are the heterotic gauge left-moving coordinates.
 Note that although the value of $k_i$ is convention dependent, the ratios of the charges of the massless 
particles $q^2/k_i$ are convention independent.  

Reference \cite{Heidenreich:2016aqi} proved a sublattice version of the WGC which applies precisely to perturbative heterotic compactifications. For a $U(1)^N$ factor, modular invariance of the worldsheet CFT leads to the existence of states with
\begin{align} q_i=l_i k_i, \quad m^2 \leq M_P^2\sum_i (g_i^2 k_i^2l_i^2),\quad l_i\in\mathbb{Z}.\end{align}
The states have charges in a sublattice with charges $(0,\ldots k_i,\ldots 0)$, and their masses satisfy eq.(\ref{wgc1}).

Consider now a $U(1)^N$ subgroup in some heterotic compactification. Applying blindly eq.(\ref{wgc1}) to this case we would expect
$N$ different new physics cut-offs $\Lambda_i^2\simeq g_i^2M_p^2$.  But in large classes of simple
heterotic string compactifications we know what the magnetic WGC scale of
new physics is, it is just the string scale, $\Lambda^2\simeq M_s^2$. That the naive magnetic bound fails is not surprising, since we already saw  that the magnetic WGC does not work in a theory which only satisfies a sublattice WGC. 

In any case, since we now what the EFT cutoff is, we can rewrite it in terms of the low energy theory data. The correct magnetic bound
corresponding to the above should  be
\beq
  \Lambda_i^2\ =\ \alpha_i k_i M_p^2  \ .
\label{modbounds}
\eeq
Unlike the original magnetic WGC, this new cutoff scale  is not motivated by any black hole considerations, but by stringy physics. Indeed, it is just the string scale! We will now make some considerations about this new cutoff scale:\begin{itemize}
\item Just like the magnetic WGC, equation (\ref{modbounds}) only works in one particular basis, the one in which charges are integer quantized. 
\item  Interestingly, the cutoff is invariant even after Higgsing.   
Our starting point is a $U(1)^N$ theory with charge lattice $\mathbb{Z}^N$. In this basis, in which the charge lattice is standard, the  $U(1)$'s are not canonically normalized. The two point functions of the corresponding CFT currents satisfy 
\begin{align}\langle J_i(x) J_j(y)\rangle\propto \frac{k_{ij}}{(x-y)^2},\label{2pgenb}\end{align}
where $k_{ij}$ is a positive-definite symmetric matrix with integer entries, the level matrix $\mathbf{k}$. Equation (\ref{2pgenb}) also tells us the 2-point function of photons in the effective field theory, which allows us to relate it to the gauge coupling matrix $\mathbf{\alpha}$ in the kinetic term
\begin{align} \frac{1}{16\pi}\int \vec{F}\wedge *(\mathbf{\alpha}^{-1}\vec{F})\end{align}
 directly via a generalization of (\ref{modbounds}),
\begin{align}\mathbf{\alpha}\mathbf{k}=16\frac{M_S^2}{M_P^2}\,\mathbf{I}.\label{xyz}\end{align}
The columns of $\mathbf{k}$ also define a sublattice of $\mathbb{Z}^N$. The sublattice WGC says precisely that this sublattice has superextremal states. 

We wish to show that (\ref{xyz}) is invariant under Higgsing. Suppose we have $J$ linearly independent fields, $J<N$, with charge vectors $\vec{v}_j$, $j=1,\ldots J$, which acquire a vev, and let $\mathbf{V}$ be the $J\times N$ matrix which has the $\vec{v}_j$ as its rows. Then, Higgsing imposes the condition $\mathbf{V}\vec{A}_{IR}=0$, so that $\vec{A}_{IR}$ lives in the kernel of $\mathbf{V}$. Let $\{\vec{n}_i\}$, $i=1,\ldots N-J$ be a orthonormal basis of $\ker\mathbf{V}$, and $\mathbf{N}$ a matrix which has these for rows. Then the charge lattice after Higgsing is the $(N-J)$-dimensional sublattice of vectors of the form
\begin{align} \Lambda_{IR}: \left\{\vec{v}\in \mathbb{R}^{N-J}: \vec{v}=\mathbf{N} \vec{q},\quad \vec{q}\in\mathbb{Z}^N\right\}.\label{lir}\end{align}
Naturally, we should also restrict the gauge coupling and the level matrix to the surviving subspace. Since $\vec{A}=\mathbf{N}^T\vec{A}_{IR}$, we get $\mathbf{k}\rightarrow \mathbf{N}^T\mathbf{k} \mathbf{N}$ and $\mathbf{\alpha}\rightarrow \mathbf{N}^T\mathbf{\alpha} \mathbf{N}$, so that (\ref{xyz}) still holds after Higgsing. 

However, we are not done yet because $\mathbf{k}\rightarrow \mathbf{N}^T\mathbf{k} \mathbf{N}$ is not really a level matrix in the same sense that $\mathbf{k}$ was in the unhiggsed theory. We defined $\mathbf{k}$ to be the charge matrix in a basis where the charge lattice is $\mathbb{Z}^{N-J}$, while $\Lambda_{IR}$ defined in (\ref{lir}) won't be in general. This can be achieved by a further general linear transformation $\mathbf{A}$, which will again preserve (\ref{xyz}).

In conclusion, it is indeed the case that  eqs.(\ref{modbounds}) and (\ref{xyz}) also apply also after Higgsing. 

\item The cutoff we obtained is secretly none other than the string scale. In any effective field theory coming from a stringy compactification, we expect the Kaluza-Klein scale $M_{KK}$ to be below or at the cutoff, so the actual cutoff of the four-dimensional effective field theory will typically be much lower than (\ref{modbounds}). 
\item As it stands now, one cannot write down the cutoff solely from effective field theory data, since the levels $k_i$ do not show up anywhere in the effective field theory; they are extra UV information. This does not mean that the cutoff is useless; as we will see below, just assuming that the theory has a heterotic stringy completion is enough to constrain the model. 
\item Unlike the WGC, we make no claim of (\ref{modbounds}) being a universal constraint which should always apply in a consistent quantum theory of gravity. Rather, it has the same range of validity as the worldsheet proof of the lattice WGC in \cite{Heidenreich:2016aqi}: It works for closed-string $U(1)$'s. 
\end{itemize}

 Let us consider now the previous $U(1)^2$ toy model embedded into
such heterotic setting.  Let us take for simplicity $g_1=g_2=g_0$ and $k_1=k_2=k_0$.  Consistency with the weakest form of the WGC is still guaranteed, since the 
particle $\psi_2$  with charge 1 under the unbroken $U(1)$ will obey the first condition  (\ref{modbounds}).The same discussion applies now as in the field theory case, and a particle $\psi_1$ with charge $(1,0)$ will couple with a strength
$\simeq g/Z$ after Higgsing. This sets the quantum of charge for $A_{IR}$, so that the charge lattice is simply $\mathbf{Z}/Z$; we get to the canonical normalization simply by multiplication by $Z$, which yields
\begin{align}k_{IR}=Z^2k_0,\quad \alpha_{IR}=\frac{\alpha_0}{Z^2}.\end{align}
  Taking this into consideration, equation (\ref{modbounds}) will predict the existence of a new physics scale at
\beq
\Lambda_s^2 \ \simeq \  k_{IR}\alpha_{IR}M_p^2=M_S.
\eeq

\section{Clockwork}

As mentioned before, the level of the current algebra $k$ is not part of the information readily available to the effective field theorist, so it is unclear how to even compute the stringy cutoff. We will now input some extra information about the spectrum, which will be enough to show that the model of \cite{saraswat} cannot be consistently embedded in the present setup, at least for parametrically large $Z$. That the Clockwork mechanism \cite{clockwork} is also somehow constrained, although the mechanism itself survives if one has a large enough number of Higgsing fields.

Let us recall that in string theory massless fields cannot have arbitrarily large charges or live in arbitrarily 
 high dimensional representations.  Massless states obey certain {\it BPS-like} conditions which strongly constrain their charges.
  The mass formula for e.g. left-movers in the string spectrum has the form (see e.g. \cite{BOOK})
 \beq
 \frac {\alpha'M^2}{4}  \ = \ h_{KM}\ +\ h_{int} \ +\ N_{osc} \ -\ 1 \ ,
\eeq
where $h_{int}$ is the contribution to the conformal weight of the state coming from the internal compact space, 
$h_{KM}$ is the contribution coming from Kac-Moody gauge degrees of freedom (from the $E_8\times E_8$ or Spin(32) lattice),
and $N_L$ counts oscilators.  Thus for a particle to be in the massless spectrum a necesary condition is to have $h_{KM}\leq 1$.
On the other hand the gauge conformal weight of a particle transforming in non-Abelian representations $R_r$ and transforming 
under Abelian groups $U(1)_s$ with charge $q_s$  and normalization $k_s$ is given by \cite{kacmoody}
\beq
h_{KM}(R_r,  k_r,k_s) \ =\  \sum_r \frac {C_r}{k_r+c_r}\ +\ \sum_s \frac {q_s^2}{k_s}\ ,
\label{hkm}
\eeq
where $k_r$ is the KM level of the associated non-Abelian, $c_r$ is the associated dual Coxeter number (e.g. $c_r=N$ for $SU(N)$),
and $k_s$ is the normalization of the Abelian factors. $C_r$ is the quadratic Casimir of the representation $R_r$.

Coming back to the $U(1)^2$ toy model, for a scalar in the representation $(Z,1)$ to be Higgsed (at least by an effective field theory mechanism), it has
to be in the string massless spectrum, hence it must have $h_{KM}\leq 1$. It  is clear from the last term in eq.(\ref{hkm}) that then  there is a bound $Z^2\leq k_0$ and
that for fixed $k_0$ the value of $Z$ cannot  be made parametrically large, one rather has
\beq 
Z^2 \ \leq k_0\ =\ \frac {16}{\alpha_0}\frac {M_S^2}{M_p^2} \ .\label{sasa}
\eeq
So in a consistent embedding of this model in a heterotic string compactification, we would get a UV cutoff
\begin{align}\Lambda^2\leq \alpha_{IR} Z^2k_0 M_P^2\leq \alpha_{IR} k_0^{2} M_P^2.\end{align}
Let us recap for a moment and see what we've got so far. As already pointed out in \cite{saraswat}, we have obtained a model by Higgsing in which the gauge coupling can be made as tiny as one wants, and in which nothing is happening at the magnetic WGC scale $\Lambda_{\text{mag.}}^2\sim \alpha_{IR} M_P^2$. Equation (\ref{modbounds}) does tell us, however, that the string scale is just a factor of $k_0$ times higher, so we do get a light cutoff assuming that $k_0$ is not too high. Equivalently, in stringy setups we cannot take $Z$ to be parametrically large, and thus get a tiny IR gauge coupling, unless $k_0\gg1$ to begin with. But (\ref{sasa}) tells us that this is equivalent to start with a tiny gauge coupling in the UV anyway, and begs the question. On top of this, there are no explicit stringy examples with parametrically large $k_0$.  We conclude that the model of \cite{saraswat} can only be consistently embedded in a heterotic compactification - or in general in the closed string sector of any compactification whatsoever - if $Z$ is not too large.

By iteration of the kind of Higgsing considered in the field theory  toy model through $N$ factors  in a $U(1)^N$ gauge group,  with equal charge $Z$ in all Higgsings,
one can obtain light fields with highly supressed charges of order $Z^{-N}$. This is the {\it clockwork} mechanism of ref.\cite{clockwork}  applied to Abelian gauge
interactions.  If we were successful in embedding such a model into the heterotic string,  the supression would be bounded as
\beq
\frac {1}{Z^N} \ \geq \ \prod_{i=1}^N \frac {1}{k_i^{1/2}}\ =\ \prod_{i=1}^N \left( \frac {\alpha_i^{1/2} M_p}{4M_S}\right) ,
\eeq
so that there is only supression if $\alpha_i \lesssim 16M_S^2/M_p^2$. Finding a specific heterotic $U(1)^N$ model with $N$ sufficiently large
and the required Higgs fields would probably be challenging though.  The bound (\ref{modbounds}) implies that the string scale is no further away than
\begin{align}\Lambda^2\leq \alpha_{IR} k_0^{2N} M_P^2.\end{align}
While the mechanism works in principle, we should also note that in the heterotic setting, $N$ is bounded from above as well. The left-moving central charge in heterotic string theory is fixed by conformal invariance to be $c_L=22$ (26 from bosonic string theory minus four of the noncompact bosons). On the other hand, a Kac-Moody algebra with group $g$ and level $k$ has a contribution to the central charge
 \begin{align}c_{KM}=\frac{k\,\text{dim}(g)}{k+c_r},\end{align}
 and in particular a $U(1)^N$ factor contributes $N$ to the central charge. It follows that, in any stringy embedding of the clockwork, $N\leq 22$\footnote{We are not counting graviphotons since these generically have Planck-suppressed couplings.}. If the model also includes the SM, the extra contribution to the central charge from the SM gauge fields is 
 \begin{align}c_{SM}=\frac{8k_{SU(3)}}{k_{SU(3)}+3}+\frac{3k_{SU(2)}}{k_{SU(2)}+1}+1\geq 4,\end{align}
which lowers the bound to $N\leq18$. In fact in generic Calabi-Yau (CY) compactifications of the heterotic string which feature no enhaced $U(1)$'s 
one rather has $N\leq 12$.

 \section{Summary}
 
Summing up, we have seen that within string theory Higgsing may produce supressed effective gauge charges  (e.g. of order $g/Z$  using $(Z,1)$ Higgsing). The mild/sublattice version of the WGC is always respected.
 Even though there is no new physics at the IR magnetic WGC scale, we have found that the string scale $M_s\sim\sqrt{k_{IR}} \Lambda_{WGC}$ is secretly lurking not too far away. Furthermore, this cutoff scale is invariant under further Higgsing. One might try to pull $k_{IR}=Z^2k_0$ up, and thus remove the low cutoff, but here is however the additional bound $Z\leq k_0$ coming from the Higgs scalar being in the string massless spectrum. Taking this into account, the string scale is necessarily $k_0 g_{IR} M_P$. Unlike $k_{IR}$, $k_0$ is directly related to details of the string compactification, so there is no natural way to make it parametrically large. In fact, we know of no explicit stringy examples with parametrically large $k_0$. For instance, $k_0\lesssim 21$ in the family of models studied in \cite{Chaudhuri:1995ve}.  
 
The same mechanism also restricts the suppression one can make in an iterative clockwork mechanism.   In this clockwork mechanism the suppression cannot be stronger than $k_0^{-N}$ where $k_0$ is the level of the UV theory. Although $k_0$ has to be small, it is still possible to achieve a large hierarchy if $N$ is large enough. So these considerations do not prevent the clockwork from being embedded into heterotic string theory. Of course, finding an actual stringy embedding of the clockwork is a much more complicated issue. In particular, it is not possible to achieve $N\gtrsim 12$ in  generic CY compactifications of the heterotic string.

\hspace{15cm}

\

\centerline{\bf \large Acknowledgments}

\bigskip

\noindent  We thank F. Marchesano, A. Uranga and I. Valenzuela for useful discussions. L. I. also thanks CERN Theory Division where part of this work
was carried out.
This work has been supported by the ERC Advanced Grant SPLE under contract ERC-2012-ADG-20120216-320421, by the grant FPA2012-32828 from the MINECO,  and the grant SEV-2012-0249 of the ``Centro de Excelencia Severo Ochoa" Programme. M.M. is supported by a postdoctoral fellowship from ITF, Utrecht University.

\end{document}